\documentclass[aps,preprint]{revtex4}
\usepackage{graphicx}
\usepackage{bm}

\begin{document}
\date{\today}
\title{Influence of branch points in the complex plane  \\
on the transmission through double quantum dots
}
\author{ I. Rotter$^1$ and A. F. Sadreev$^{2,3}$}
\affiliation {$^1$ Max-Planck-Institute f\"ur Physik komplexer
Systeme, D-01187 Dresden, Germany }
\affiliation{$^2$  Kirensky
Institute of Physics, 660036, Krasnoyarsk, Russia}
\affiliation{$^3$ Department of Physics and Measurement,\\
Technology Link\"{o}ping  University,  S-581 83 Link\"{o}ping,
Sweden}

\begin{abstract}
We consider  single-channel
transmission through a double quantum dot system consisting of
two single dots that are connected by a wire and coupled
each to one  lead. The system is described
in the framework of the $S$ matrix theory by using the effective Hamiltonian
of the open quantum system. It
consists of the Hamiltonian of the closed system
(without attached leads) and a term that accounts for the coupling
of the  states via the
continuum of propagating modes in the leads. This model allows to study
the physical meaning of branch points in the complex plane.
They are points of coalesced  eigenvalues and
separate the two scenarios with avoided level crossings
and without any crossings in the complex plane.
They influence strongly the features of transmission through double quantum
dots.

\end{abstract}
\maketitle

\section{Introduction}

The phenomenon of avoided level crossing (Landau-Zener effect) is
studied theoretically as well as experimentally for many years. It is a
general property of the discrete states of a quantum system the energies
of which will never cross when the interaction between them is  nonvanishing.
Their wave functions are
exchanged at the critical value of  a certain tuning parameter
where the avoided crossing takes place.
The reason for the avoided crossing of two discrete levels follows from the
expression  for the two
eigenvalues $e_\pm$ of the Hamiltonian of the system,
\begin{eqnarray}
e_{\pm}= \frac{e_1+e_2}{2}\pm\frac{1}{2}\sqrt{(e_1-e_2)^2+4 \omega^2}
\nonumber
\label{discr}
\end{eqnarray}
where $e_1$ and $e_2$ are the energies of the non-interacting states
and $\omega$
is their interaction. Since the square root contains only positive values,
$e_+$ and $e_-$ are always different from one another with the only exception
of vanishing interaction $\omega$.

A crossing point of the two eigenvalues $e_\pm$
can be found by continuing into the
complex plane, i.e. by adding a negative term into the square root. The
mathematical properties of such a crossing point in the complex plane
are discussed in many
papers. According to Kato \cite{kato}, they are called exceptional points,
since the spectrum is supposed to be incomplete at these points.
The exceptional points are branch points in the complex plane
\cite{moisrep,ro01}.
Although the number of these points in the complex plane is of measure zero,
their meaning for physical processes is large. They are
related to the phenomenon of avoided crossing of discrete
states as shown already in \cite{kato}.

In recent studies, it turned out  that not only discrete states
avoid crossing, but  also resonance states do not cross, as a rule
\cite{ro01,dicaro,mudiisro,marost,rep}.
An avoided level crossing in the complex
plane is accompanied by a redistribution of the spectroscopic properties of
the resonance states. Most interesting is the bifuraction of decay widths
related to the avoided  crossing of levels in the complex plane since it
causes different time scales in the system. The long-lived (trapped) resonance
states cause narrow resonances in the cross section on a
weakly energy dependent  background  induced by  the short-lived resonance
states. A similar situation is discussed recently in \cite{mois}.
The resonance trapping phenomenon discussed in
\cite{ro01,dicaro,mudiisro,marost,rep}
is a collective coherent resonance phenomenon as stated also in \cite{mois}.
The  avoided level crossings may form a branch cut \cite{mois}. This cut
can be traced up to the avoided crossing of discrete states \cite{ro01}.

Often, the branch points in the complex plane are
identified with double poles of the $S$ matrix \cite{newton}
when related to physical processes. It became possible directly to study
the spectra of atoms in the very neighborhood of double poles of the $S$
matrix  by means of laser fields \cite{marost,atom1,atom2,atom3}.
The results show a smooth
behavior of the observables when crossing the double pole by tuning
the parameters of the laser field. Moreover, recent studies in the
framework of  schematical models have
shown that the branch points in the complex plane
separate the scenario with
avoided level crossing from that without any crossing  \cite{ro01,vanroose}.
In \cite{ro01,dicaro,mudiisro,marost,rep}, the double poles
of the $S$ matrix are identified
with points at which the eigenvalues of two states of the effective
Hamiltonian coalesce. In \cite{mudiisro,marost,thomas,marost4}, the
line shape of
resonances in the very neighborhood of double poles of the $S$ matrix is
studied.

In \cite{1} the $S$ matrix theory is applied to the transmission
through double
quantum dots (QDs) consisting of two single QDs and a wire connecting them.
The study of these artificial molecules is of great interest since they
display the simplest structures of quantum-computing devices that can be
controlled by external parameters, e.g. \cite{qdot1,qdot2}.
One of the interesting results obtained for a double QD system,
is the appearance of transmission zeros of different order
at energies that are related to the eigenvalues of the Hamiltonians of the
single QDs \cite{1}. They appear even in cases when the transmission is
large in this  energy region. In such a case, they can be seen as narrow
dips in the transmission probability.

Double dot systems provide a very powerful tool for studying the properties
of branch points in the complex plane and their physical meaning.
When leads are attached to them, the double dot systems
allow further to study the relation of the branch points in the complex
plane to both the double poles of the $S$ matrix and the
points  where two eigenvalues of the Hamiltonian of the open
quantum system coalesce. This is, above all, due
to the symmetries involved in the system in a natural manner. Moreover,
the properties of a double dot system  can be controlled by
external parameters in a very clear manner. The double QD itself is
characterized by the coupling strengths $u$ between the wire and
the single QDs, the spectral properties of the two single QDs,
as well as by the length  and the width  of the wire.
The coupling of the double dot system to the environment is given by the
coupling strength $v$ to the leads attached to it. All these parameters are
well defined and can be controlled. One may call $v$ the external
coupling of the double QD system (via the leads) and $u$ the internal coupling
(via the wire) that is characteristic of the double dot system as a whole.

In the present paper, we will study  a simple model for a
double QD system
with the aim to receive information on the branch points in the complex
plane and their relation to physical processes such as transmission.
We use $S$ matrix theory combined with
the method of the effective Hamiltonian which consists of
two parts. The first part is the (Hermitian) Hamiltonian of the closed system
and the second part is an additional (non-Hermitian) term that
takes into account the coupling of the  states of the system via
the continuum. The continuum is given by the modes propagating in the
two half-infinite 1d-leads
when attached to the system. The interplay between these two parts of the
effective Hamiltonian characterizes the different physical situations.

In Sect. II, we give the $S$ matrix for the transmission through a
model double QD system by using the effective Hamiltonian formalism.
The double QD consists of two single QDs with one  state
in each, a wire with a single eigenenergy that depends on  the
length of the wire, and with one channel for the propagation
of the mode in the attached leads. We define the
spectroscopic values $E_k$ and $\Gamma_k$ of the resonance states $k$.
In Section III, we study analytically the features of the
eigenvalues and eigenvectors at
the branch point in the complex plane. Here, at a certain energy $E=E_c$,
two eigenvalues   of the effective Hamiltonian  coalesce.
We show numerical examples obtained for
branch points in the complex plane as well as for the transmission through the
double dot system. The branch points can be seen
by varying different parameters.
The transmission scenario at small $v/u$ is characterized by
transmission peaks which
are spread over a certain energy region that is the larger the
larger the internal interaction $u$ is.
In contrast to this picture, the transmission peaks
are no longer spread in energy when  $v/u$ is large.
Here, level attraction and width bifuraction take place with the
consequence that one  narrow resonance appears on a
smooth background created by the two  broad resonance states.
The separation between the two different
scenarios is provided by the branch point in the complex plane.
This separation is independently of whether the eigenstates cross
or avoid crossing in the complex plane at the energy $E_c$.

In Sect. IV, we consider the effective Hamiltonian as well as the transmission
through the double dot system when it is coupled with different strength
to the two leads. In the following section V,   we show
numerical examples for  transmission and  eigenvalue trajectories
of a double dot system with altogether
five and eleven, respectively,  states as a function of both,
the length  of the wire and the (external) coupling strength $v$.
The main features of the eigenvalue trajectories as well as of the
transmission are the same as those discussed in Sect. III.
Moreover, we draw some conclusions on the different bonds of the two single
QDs in the artificial molecule. The appearance of  different bond
types is also related to the positions of
the branch points in the complex plane. In the last section, we summarize the
results obtained.


\section{Effective Hamiltonian and $S$ marix theory for transmission through
coupled quantum dots}

In our study, we follow the paper \cite{saro} where the $S$ matrix theory
for transmission
through QDs in the tight-binding approach is formulated, and
the paper \cite{1} where the $S$ matrix theory is applied to a double QD
system consisting of two single QDs coupled to each other by a wire.
As in \cite{1},  we consider a simple model with   a small number of states
in each single QD and one mode propagating  through the wire. This simple
model is able to explain  the characteristic features of the
transmission  through realistic double dot systems of the same structure, as
shown in  \cite{1}.

First we will consider the simplest case
with  only one state $\varepsilon_1$ in each single dot and
one mode $\epsilon(L)$  propagating in the wire of length $L$.
The wire and the single QDs are coupled by $u$.
The effective Hamiltonian of such a system is \cite{saro,1}
\begin{equation}
\label{Heffgen}
H_{\rm eff}=H_B+\sum_{C=L,R}
V_{BC}\frac{1}{E^{+}-H_C}V_{CB},
\end{equation}
where
\begin{equation}\label{HB3}
  H_B=\left(\begin{array}{ccc} \varepsilon_1    & u & 0 \cr
         u & \epsilon(L) & u \cr
            0  &  u &  \varepsilon_1  \end{array}\right)
\end{equation}
is the Hamiltonian of the closed double dot system, $H_C$ is the
Hamiltonian of the left  ($C=L$) and right ($C=R$) reservoir and
$E^{+}=E+i0$. The second term of $H_{\rm eff}$ takes into account
the coupling of the eigenstates of $H_B$ via the reservoirs when
the system is opened. It introduces  correlations between the
states of an open quantum system that appear additionally to those
of the closed system  \cite{rep}. The effective Hamiltonian
$H_{\rm eff}$ is non-Hermitian.

The coupling matrix between the closed double dot system and
the reservoirs can be found if both are specified. We take the
reservoirs (leads) as semi infinite one-dimensional wires in
tight- binding approach \cite{saro}. The connection points
of the coupling between the system and the reservoirs are at the edges of the
one-dimensional leads. Then the coupling matrix elements take the
following form \cite{saro,1}
\begin{eqnarray}
\label{Vm} V_m(E,L)=v\psi_{E,L}(x_L)\psi_m(j=1)=v\sqrt{\frac{\sin
k}{2\pi}}\psi_m(1),\nonumber\\
V_m(E,R)=v\psi_{E,L}(x_R)\psi_m(j=3)=v\sqrt{\frac{\sin k}{2\pi
}}\psi_m(3),
\end{eqnarray}
where $k$ is the wave vector related to the energy by $E=-2\cos k$,
~$\psi_m(j), j=1, 2, 3$, are the eigenfunctions of (\ref{HB3}), and
$v$ is the hopping matrix element between the edge of the lead
and the QD. The $v$ will be varied in our calculations.
The eigenvalues of the Hamiltonian (\ref{HB3}) are real,
\begin{eqnarray}\label{ED3}
  E_{1,3}^B &=&\frac{\varepsilon_1+\epsilon(L)}{2} \mp \eta,
\quad  E_2^B = \varepsilon_1,
  \\
&&\quad {\rm with} \quad
  \eta^2=\Delta\varepsilon^2 + 2u^2, \; \;\Delta\varepsilon=
  \frac{\varepsilon_1 - \epsilon(L)}{2} \; ,
\end{eqnarray}
and the eigenstates read
\begin{equation}\label{psiB3}
  |1\rangle =\frac{1}{\sqrt{2\eta(\eta+\Delta\varepsilon)}}\left(\begin{array}{c} -u\cr
           \eta+\Delta\varepsilon \cr
            -u \end{array}\right), \;
            |2\rangle =\frac{1}{\sqrt{2}}\left(\begin{array}{c} 1 \cr
           0 \cr    -1 \end{array}\right), \;
  |3\rangle =\frac{1}{\sqrt{2\eta(\eta-\Delta\varepsilon)}}\left(\begin{array}{c} u\cr
           \eta-\Delta\varepsilon \cr
            u \end{array}\right).
      \end{equation}
As a result, we get the following expression for the
effective Hamiltonian \cite{1},
\begin{equation}
\label{Heff3}
H_{\rm eff}=\left(\begin{array}{ccc}
E_1^B-\frac{v^2u^2e^{ik}}{\eta(\eta+\Delta\varepsilon)}& 0 &
\frac{v^2ue^{ik}}{\sqrt{2}\eta} \cr 0& \varepsilon_1-v^2e^{ik} & 0
\cr  \frac{v^2ue^{ik}}{\sqrt{2}\eta} & 0 &
E_3^B-\frac{v^2u^2e^{ik}}{\eta(\eta-\Delta\varepsilon)}\cr
             \end{array}\right) \; ,
\end{equation}
which is symmetric. Its complex
eigenvalues $z_k$ and eigenvectors $|k)$  are \cite{1}
\begin{eqnarray}
\label{poles3}
z_2 & = & \varepsilon_1-v^2e^{ik},\nonumber\\
z_{1,3} & = & \frac{\varepsilon_1+\epsilon(L)-v^2e^{ik}}{2} \mp
  \sqrt{\left(\frac{\epsilon(L)-\varepsilon_1+v^2e^{ik}}{2}\right)^2+2u^2}
\end{eqnarray}
and
\begin{eqnarray}
|1)=\left(\begin{array}{c}
  a \cr   0\cr   b\end{array}\right)\, , \qquad
|2)=\left(\begin{array}{c} 0 \cr
           1 \cr    0 \end{array}\right) \, , \qquad
|3)=\left(\begin{array}{c}b \cr 0\cr -a
             \end{array}\right) \, ,
\label{psiHeff3}
\end{eqnarray}
 where
\begin{eqnarray}
\label{Delxi}
  a & = & -\frac{f}{\sqrt{2\xi(\xi+\omega)}}\; ,\;\qquad \;
    b=\sqrt{\frac{\xi+\omega}{2\xi}}  \nonumber \\[.2cm]
  f & = & \frac{v^2ue^{ik}}{\sqrt{2}\eta}\, , \quad
  \omega=-\eta+\frac{\Delta\varepsilon v^2e^{ik}}{2\eta}\, , \quad
  \xi^2=\omega^2+f^2.
\end{eqnarray}
The eigenfunctions are biorthogonal, $H_{\rm eff} |k) = z_k |k)$ with
 \cite{rep}
\begin{eqnarray}
(k|l) \equiv \langle k^* | l \rangle = \delta_{k,l} \, .
\label{biorth}
\end{eqnarray}
Using the eigenvalues (\ref{poles3}) and
eigenfunctions (\ref{psiHeff3})  of the effective
Hamiltonian, the amplitude for the transmission
through the double QD takes  the simple form \cite{saro}
\begin{equation}
\label{trHeff3}
t=-2\pi
i\sum_{\lambda}\frac{\langle L|V|\lambda)(\lambda|V|R\rangle }{E-z_{\lambda}}.
\end{equation}
Substituting (\ref{Vm}), (\ref{psiB3}) and (\ref{psiHeff3})
into the  matrix elements $\langle L|V|\lambda)$ and
$(\lambda|V|R\rangle $ we obtain
\begin{eqnarray}\label{Vlam}
\langle L|V|2)&=&\sum_m\langle E,L|V|m\rangle \langle
m|2)=\frac{v}{2}\sqrt{\frac{\sin k}{\pi}},\nonumber\\
(2|V|R\rangle &=&\sum_m(2|m\rangle \langle m|V|E,L\rangle
=-\frac{v}{2}\sqrt{\frac{\sin k}{\pi}},\nonumber\\ \langle
L|V|1)=(1|V|R\rangle & =& v\sqrt{\frac{\sin
k}{2\pi}}(\psi_1(1)a+\psi_3(1)b), \nonumber\\ \langle
L|V|3)=(3|V|R\rangle &=&v\sqrt{\frac{\sin
k}{2\pi}}(\psi_1(1)b-\psi_3(1)a).
\end{eqnarray}
The transmission probability is $T=|t|^2$.

The spectroscopic values such as the positions in energy
of  states are originally defined for the discrete eigenstates
of  Hermitian Hamilton operators that describe  closed quantum systems.
The  decay widths do not appear explicitely
in this formalism since the eigenvalues
of the Hamiltonian are real. They are calculated from the tunneling matrix
elements  by means of the eigenfunctions of this Hamiltonian.
The corresponding values for resonance states are energy dependent functions
since the eigenvalues as well as the eigenfunctions of the non-Hermitian
effective Hamilton operator (\ref{Heffgen}) depend on energy.
Nevertheless, spectroscopic values for resonance states can be defined,
and that by solving the fixed-point equations \cite{rep}
\begin{eqnarray}
\label{fixed1}
E_k & = &
{Re}\big(z_k\big)\big|_{E=E_k}
\end{eqnarray}
and defining
\begin{eqnarray}
\label{fixed2}
\Gamma_k & = &
2 \, {Im}\big(z_k\big)\big|_{E=E_k} \, .
\end{eqnarray}
The values $E_k$ and $\Gamma_k$ characterize a resonance state
whose position in energy  is $E_k$ and whose decay width
is $\Gamma_k$. This resonance state causes a resonance
of Breit-Wigner type in the cross section when it is well separated from other
resonance states. In the regime of overlapping  resonances, the
relation between $E_k$ and $\Gamma_k$ on the one hand, and the resonances
seen  in the cross section on the other hand, is less well defined.

In the denominator of the $S$ matrix, the eigenvalues $z_k$ of
the effective Hamiltonian $H_{\rm eff}$ appear  in their full
energy dependence. That means that at every energy $E$ of the
system,  the contribution of every resonance state $k$ is taken
into account in correspondence to the value $z_k(E)$. This fact
becomes important when $z_k(E_{l\ne k}) \ne z_k(E_k)$ and the
contribution of the resonance state $k$ can not be neglected at
the energy $E=E_l$, i.e. when the resonance states overlap.

Another definition of the spectroscopic values of a resonance state is
by means of the poles of the $S$ matrix. This (standard)
definition of the spectroscopic values in the framework of the $S$ matrix
theory is not a direct one since the poles of the $S$ matrix
give information on the resonances, but not on the spectroscopic
properties of the resonance states.  The $S$ matrix has a pole
only  when the energy is continued into the complex plane.
We remind however that the $S$ matrix describing physical processes is defined
for real energies $E$, and $|S|^2 \le 1$. It is not surprisingly therefore
that the two definitions do not coincide completely.
In the following, we will characterize the resonance states by the energy
dependent eigenvalues $z_k$ and eigenfunctions $|k)$ of the effective
Hamiltonian $H_{\rm eff}$ as well as by the values $E_k$ and $\Gamma_k$,
but not by the poles of the $S$ matrix. The reason for doing this, is
the clear definition of the spectroscopic values $E_k$ and $\Gamma_k$
also in the regime of overlapping resonances \cite{rep}, by means of the
effective Hamiltonian $H_{\rm eff}$ that describes the open quantum system.

It may happen that, at a certain point, $z_k=z_l$
for two different states $k$ and $l$.
Such a  point might be considered as the analogue of a double pole of
the $S$ matrix. However,  the coalescence of two eigenvalues
$z_k, ~z_l$ at a certain energy $E_c$ does not mean that also
the poles exactly coincide.
Therefore, we will not consider double poles of the  $S$ matrix in the
following, but will look at the points and their energies $E_c$
where  the two eigenvalues $z_k, \, z_l$ coalesce. In such a case, the
transmission is determined mainly  by interferences between the two
resonance states $k$ and $l$. These interferences
influence strongly the line shape of resonances \cite{rep,marost4}.

Generally, two resonance states $k$ and $l$
avoid crossing in the complex plane, i.e. the eigenvalues
$z_k$ and $z_l$ coalesce at an energy $E=E_c $ that is different from the
energies $ E_k,\, E_l$.
The phenomenon of avoided  crossing of resonance states in the complex
plane is in  complete analogy
to the well-known phenomenon of  avoided  crossing of discrete states.
In the latter case, the crossing point can be found  by opening the system
and varying the coupling strength of the discrete states to the continuum,
i.e. by continuing into the complex plane.
In both cases, the crossing point influences strongly the properties of
the states although it is hidden \cite{ro01}.

The formalism for the description of double QDs with more complicated
structure is given in \cite{1}. We will not repeat it here. We will
however use it to obtain
some numerical results for the transmission through double QDs
with a larger number of states.

\section{Branch points in the complex plane}

Let us define the value
\begin{eqnarray}
F= \left(\frac{\epsilon(L)-\varepsilon_1+v^2e^{ik}}{2}\right)^2+2u^2
\label{Fdef}
\end{eqnarray}
by which the two eigenvalues  $z_{1,3}$ differ
according to (\ref{poles3}). $F$ is real only when $k= n \pi; ~n=0, 1, ..$.
When $F>0$,  Eq.  (\ref{poles3}) gives repulsion of the two levels
1 and 3 in their energies   $Re(z_k)$. When however $F<0$, there is a
bifurcation of the decay widths $Im(z_k$).

Most interesting is the case  $F=0$ since the eigenvalues and eigenvectors
of $H_{\rm eff}$ have some special properties under this condition.
From (\ref{poles3}) follows   $z_1=z_3$ for the eigenvalues,
i.e. the condition $F=0$ defines a point of coalesced eigenvalues.
According to (\ref{psiHeff3}), the  components of the
(complex) eigenvectors $|1) $ and $|3)$ become infinitely large, and
\begin{eqnarray}
|1) = \pm \; i\; |3) \qquad {\rm when} \quad F=0 \, .
\label{coal}
\end{eqnarray}
Also the normalization condition (\ref{biorth}) is fulfilled when $F=0$
due to the biorthogonality of the eigenfunctions, since the difference
between two infinitely large numbers may be 0 or 1.
These relations between the eigenvalues and eigenvectors of $H_{\rm eff}$
that follow from the condition $F=0$, hold not only
for the special case considered here. They hold also for the eigenvectors
of an effective Hamiltonian that describes atoms under the influence of a
laser field \cite{marost}. More generally, they
characterize  the eigenstates of an effective Hamiltonian  that
describes an open quantum system  \cite{ro01,rep,korsch}.

The point at which $F=0$, is a branch point in the complex plane
\cite{ro01,moisrep,rep}. This point separates the
scenarios with level repulsion on the one hand and width
bifurcation on the other hand \cite{ro01,rep}. The study on the basis of
a schematical model provided the following additional results:
level repulsion is
accompanied by the tendency to reduce the differences between the widths
of the two states, while width bifurcation is accompanied by level clustering.

According to (\ref{poles3}), the two eigenvalues $z_1$ and $z_3$ of the
effective Hamiltonian (\ref{Heff3}) coalesce
when $Re(F)=0$ and $Im(F)=0$. The first condition gives
\begin{equation}
\label{branch} v_c^4 = (\epsilon(L_c) - \varepsilon_1^c)^2+8u_c^2.
\end{equation}
From the second condition and $E=-2cos(k)$, we find the
energy at which the coalescence takes place,
\begin{equation}\label{Ebranch}
E_c =\frac{2\, (\epsilon (L_c)-\varepsilon_1^c)}{v_c^2}.
\end{equation}

\begin{figure}[t]
\includegraphics[width=.5\textwidth]{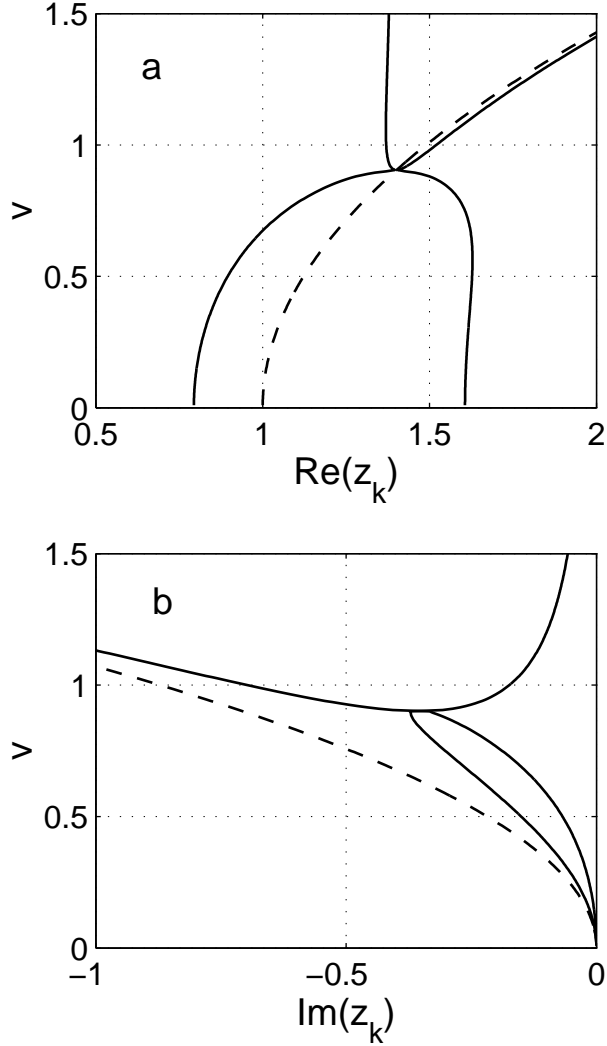}
\caption{The evolution of $Re(z_k)$ (a) and $Im(z_k)$ (b), ~$k=1, 3$
(solid lines), ~$k=2$ (dashed line),  of the three eigenvalues of the
effective Hamiltonian $H_{\rm eff}$ as a function of $v$ at
$E=E_c=0.9847$. The parameters of the double DQ system are chosen
as $\varepsilon_1=1, ~\epsilon(L)=2-L/5, ~u=1/4,
~L=3$. At $v=v_c=0.9013$ the two eigenvalues $z_1$ and $z_3$
coalesce. The $Re(z_1)$ and $Re(z_3)$ approach each
other when $v<v_c$, while the corresponding $Im(z_1)$ and $Im(z_3)$
bifurcate when  $v>v_c$. At the branch point in the complex plane
$E_c\ne E_k, ~E_l$.
}
\label{fig1}
\end{figure}

\begin{figure}[t]
\includegraphics[width=.5\textwidth]{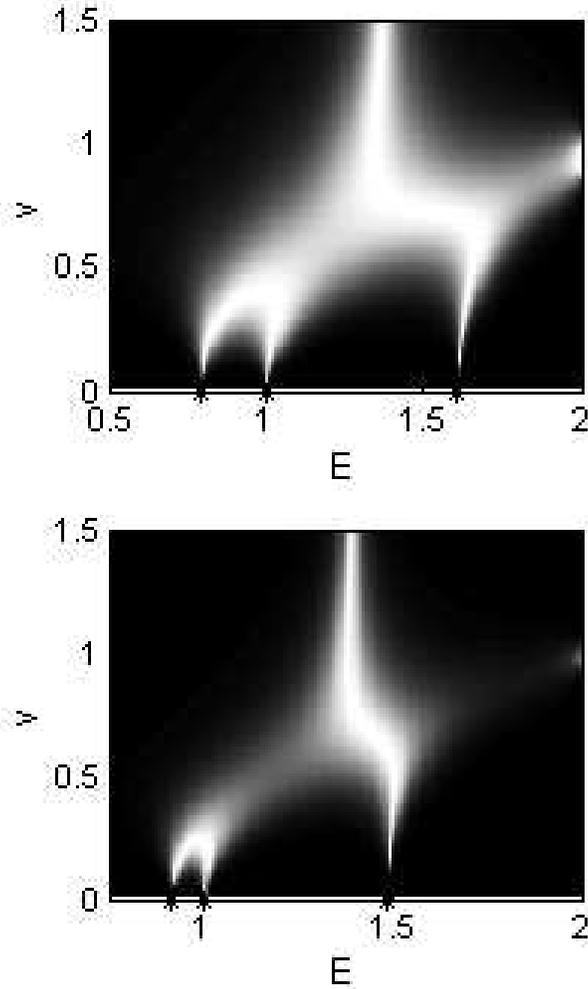}
\caption{ The transmission probability through the double QD
versus $v$ and energy. Each single QD has one level at
$\varepsilon_1=1$. It is $\epsilon(L)=2-L/5$ and $L=3$.
The eigenenergies of the double QD are shown by stars.  The
case (a) corresponds to Fig. \ref{fig1}.
The coupling constant between the single dots and the wire is $u=1/4$.
The point of coalesced eigenvalues
is $v_c=0.9013, ~E_c=0.9847$, and the solutions of the fixed point
equations (\ref{fixed1})  give $E_k=E_l\ne E_c$ as can be seen from
Fig. \ref{fig1}.
In the case (b), the
coupling constant $u=u_c=0.1443$ between the single dots and the wire
is chosen in correspondence to Eq. (\ref{ufix}). Therefore,
 $E_k =E_l$ coincides with  $E_c=7/5$ at   $ v=v_c$.
}
\label{fig2}
\end{figure}

In Fig. \ref{fig1}, we present the
typical evolution of the real and imaginary parts of the eigenvalues $z_k$
of the effective Hamiltonian $H_{\rm eff}$ as a
function of the  coupling constant $v$.
The parameters of the system are $\epsilon=2-L/5, \; \varepsilon_1=1,
\; u=0.25, \; L=3$. With these parameters, it follows from Eqs. (\ref{branch})
and (\ref{Ebranch}) that the eigenvalues $z_{1,3}$ coalesce when
$v=v_c=(1/2+9/25)^{1/4}= 0.9013$  and $ E=E_c=0.9847$.
The results shown in Fig. \ref{fig1} are obtained for the energy
$E=E_c$. Although there are three eigenstates, only $z_1$ and $z_3$
coalesce at  the point $E=E_c, ~v=v_c$. The second eigenstate
does not interact with the two other ones since it is not directly
coupled to the leads. It is coupled to the leads only via the two
single QDs, and this coupling is symmetrically. This result is in
accordance  to (\ref{Heff3}).
It can be seen further, that the two states $|1)$ and $|3)$
with energies $Re (z_k), \; Re (z_l)$ coalesce (when $v=v_c$) at
the energy $E=E_c$. At this branch point in the complex plane
$E_k  \ne Re (z_k)|_{E=E_c}, \; E_l \ne Re (z_l)|_{E=E_c}\; .$
This means, the two resonance states $|1)$ and $|3)$ do cross
at $E=E_c$ but not at the energy $E_k$ or $E_l$.
In Fig. \ref{fig2} (a), the corresponding transmission probability
versus $v$ and $E$  is shown.

Let us consider now the behavior of the eigenvalues of the
effective Hamiltonian as a function of $v$
at the energy $E=E_k$ where $E_k=Re(z_k(E_k))$
is  solution of Eq. (\ref{fixed1}). In the general case,
it is not easy to find the solution of the fixed point
equation. However for the energy
(\ref{Ebranch}) at which the eigenvalues $z_k$  coalesce, Eq.
(\ref{fixed1}) can be easily solved analytically. From (\ref{poles3}),
(\ref{branch}) and (\ref{Ebranch}) we obtain
\begin{equation} \label{fixE}
  E_k=\epsilon(L_c)=\frac{2(\epsilon(L_c)-\varepsilon_1^c)}{v_c^2}
\end{equation}
and
\begin{equation} \label{ufix}
u_c^2=\frac{(\epsilon(L_c)-\varepsilon_1^c)^2}{8}\left(
\frac{4}{\epsilon(L_c)^2}-1\right).
\end{equation}
With the parameters chosen in Fig. \ref{fig1}, the
last equation implies that solutions exist if $\epsilon(L)\leq 2$. We can
consider therefore the evolution of the eigenvalues  $z_k$ with $v$
at $E=E_k=\epsilon(L)=7/5$ and look for the point where the two eigenvalues
coalesce. The critical values at the branch point in the complex plane
are $ u_c=0.1443$ and  $E_c=7/5$. The
evolution of the eigenvalues  $z_k$ with $v$ for $u_c, ~E_c, ~L=3$
is  similar to that given in  Fig. \ref{fig1}. It is not shown here.
The corresponding transmission picture Fig. \ref{fig2} (b)
is also similar to  Fig. \ref{fig2} (a).
The main difference is the smaller spreading of the
eigenvalues of $H_B$ and the smaller transmission probability
according to the smaller value $u$ in the case with $E_k=E_l=E_c$.
In both cases, the transmission is more spread in energy
at $v<v_c$  than at $v \ge v_c$. This is in accordance with level
repulsion seen in the eigenvalue trajectories at small $v$ and
level attraction appearing  at large $v$. There is a
transmission peak  at $v\approx 1$ near the upper border $E = 2$
in both cases. This peak follows from the energy dependence of the $Re(z_k)$:
the positions of  the two resonance states with large width
approach  $E=2$ with $v\approx 1$ (see Fig. \ref{fig1} where the eigenvalues
are shown for an energy $E<2$).  We can state therefore
that the characteristic features of the transmission pictures
do not depend on whether
the two states avoid crossing or cross in the complex plane.

\begin{figure}[t]
\includegraphics[width=.5\textwidth]{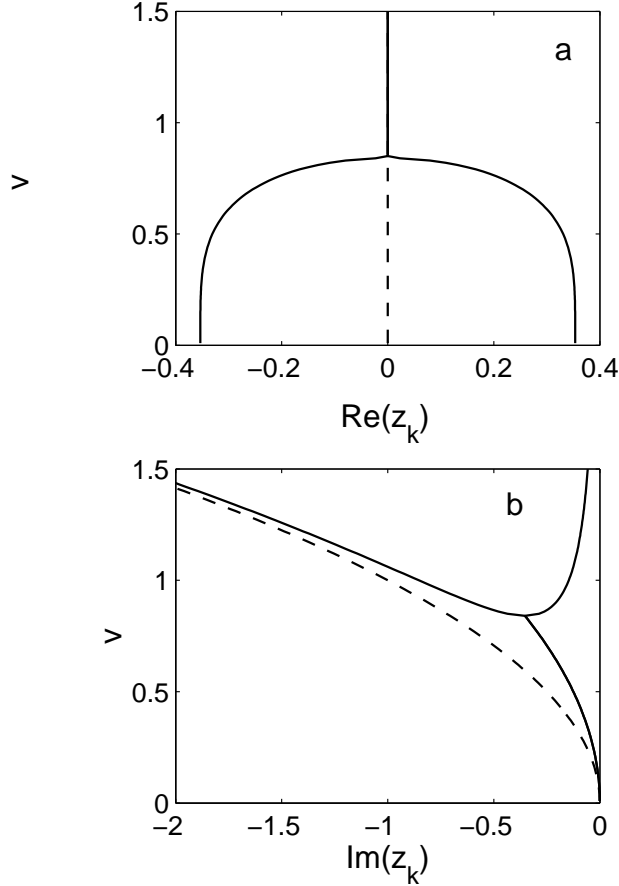}
\caption{The evolution of $Re(z_k)$ (a) and $Im(z_k)$ (b), ~$k=1, 3$
(solid lines), ~$k=2$ (dashed line),
of the three eigenvalues of the effective Hamiltonian $H_{\rm
eff}$ as a function of $v$ at $E=E_c=0$.  The parameters $u=1/4, ~L=10,
~\varepsilon_1=0, ~\epsilon(L)=2-L/5$  of the double QD system
are chosen in such a manner that $\epsilon(L)=\varepsilon_1=0$ at
$E=0$. Here, the two eigenvalues coalesce. ~$v_c=8^{1/4}u^{1/2}=0.8409$.  }
\label{fig3}
\end{figure}

\begin{figure}[t]
\includegraphics[width=.5\textwidth]{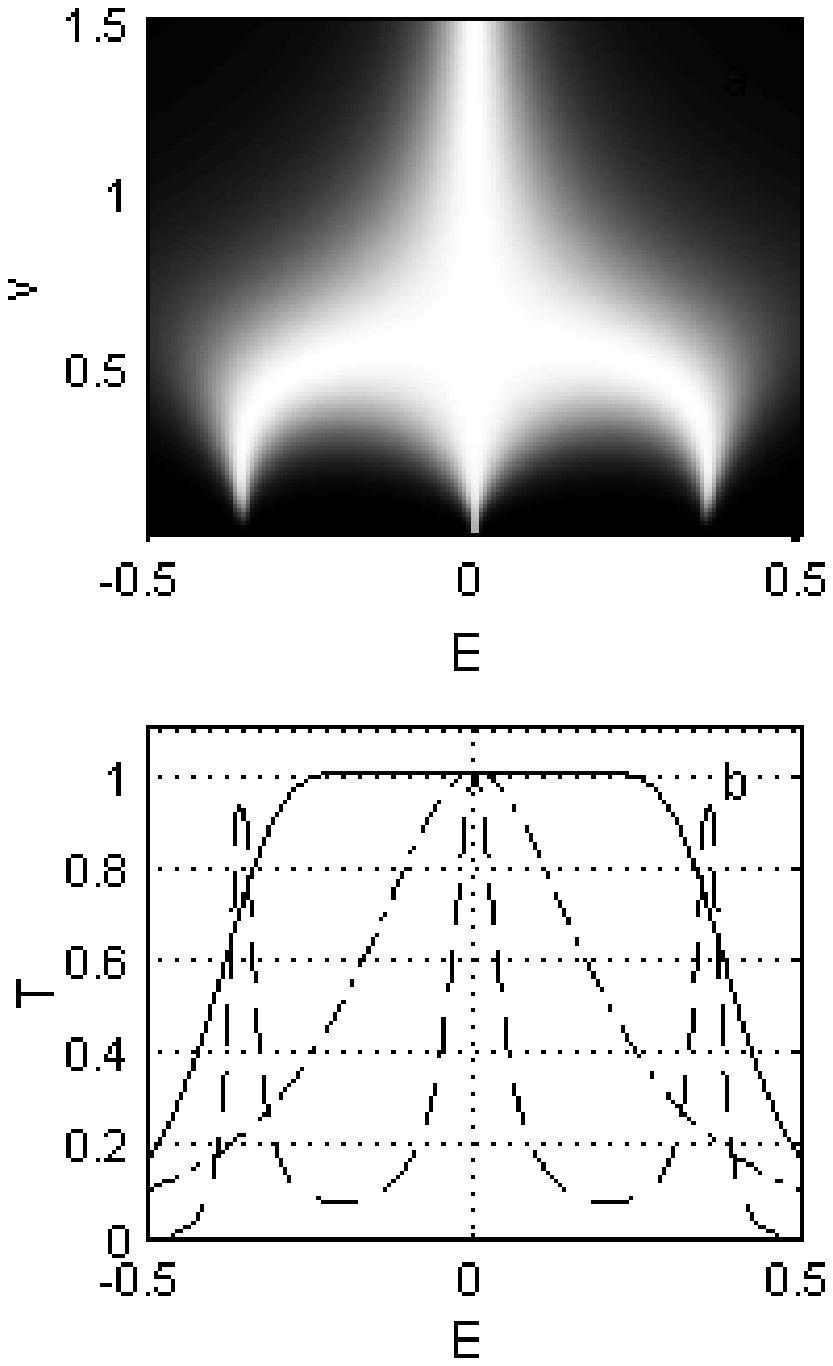}
\caption{(a) The transmission probability through the double QD
versus $v$ and energy for the case shown in Fig. \ref{fig3}. (b)
The same as (a) but for fixed $v=0.2$ (dashed line), $v=0.53$
(solid line), and $v=0.83$ (dot-dashed line).
At $v=0.53$, the double QD is a perfect filter.
}
\label{fig4}
\end{figure}

In Fig. \ref{fig3}, we present the  peculiar symmetrical
behavior of the eigenvalues $z_k$  versus $v$ at $E=0$ for
the resonant case with the parameters $\epsilon(L)=\varepsilon_1, \; L=5$.
In this case  we have, according to  Eqs. (\ref{branch})
and (\ref{Ebranch}), $E_c=0$ and $v_c= 8^{1/4}u^{1/2}$.
At $v<v_c$, the widths of the two states 1 and 3 are equal,
$Im(z_1)=Im(z_3)$, while at $v>v_c$ their positions are equal,
$Re(z_1)=Re(z_3)$. The state 2
is not involved in the crossing scenario as in Fig. \ref{fig1}.

The transmission probability versus energy and $v$ is presented in
Fig. \ref{fig4}. It has the same symmetrical behavior as the eigenvalue
pictures. Of special interest is, as Fig. \ref{fig4} (b) shows,
that this symmetrical case is at $v=0.53$ a
perfect filter: the transmission probability is equal to one in a large
energy range.

\begin{figure}[t]
\includegraphics[width=.6\textwidth]{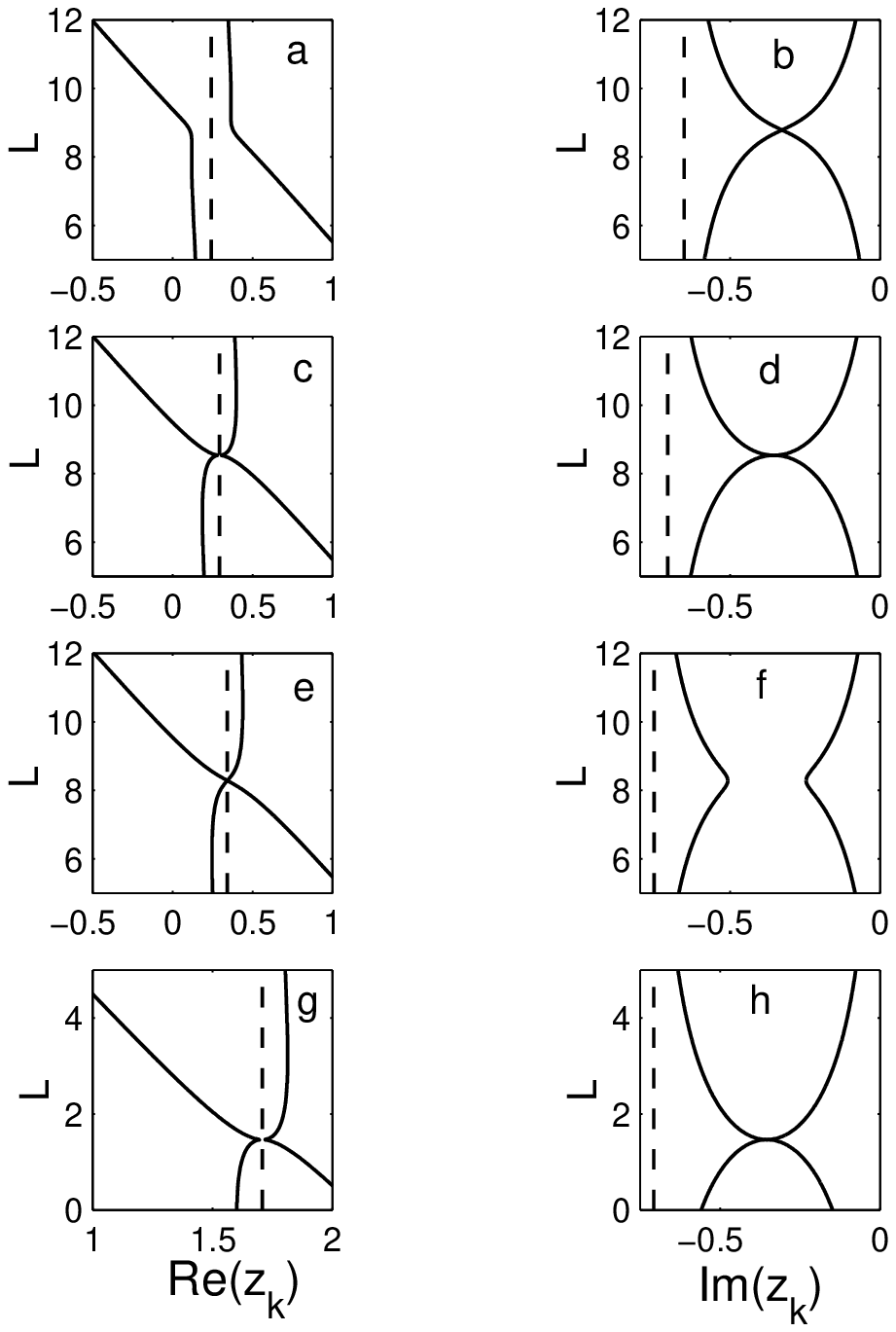}
\caption{The evolution of the real and imaginary parts of the eigenvalues
$z_k$,  ~$k=1, 3$ (solid lines), ~$k=2$ (dashed line),
as a function of the length $L$ for
the same double QD system as in Fig. \ref{fig1} but $v=1$.
$E_c=\pm \sqrt{2}$.
The critical values of the length $L$ at the two points of coalescence
of eigenvalues are $L_{1c}=1.4645, ~L_{2c}=8.5355$.
(a, b) $E=-\sqrt{2}-0.1$, ~(c, d) $E=-\sqrt{2}$,
~(e, f) $E=-\sqrt{2}+0.1$, and (g, h) $E=\sqrt{2}$.} \label{fig5}
\end{figure}

\begin{figure}[t]
\includegraphics[width=.7\textwidth]{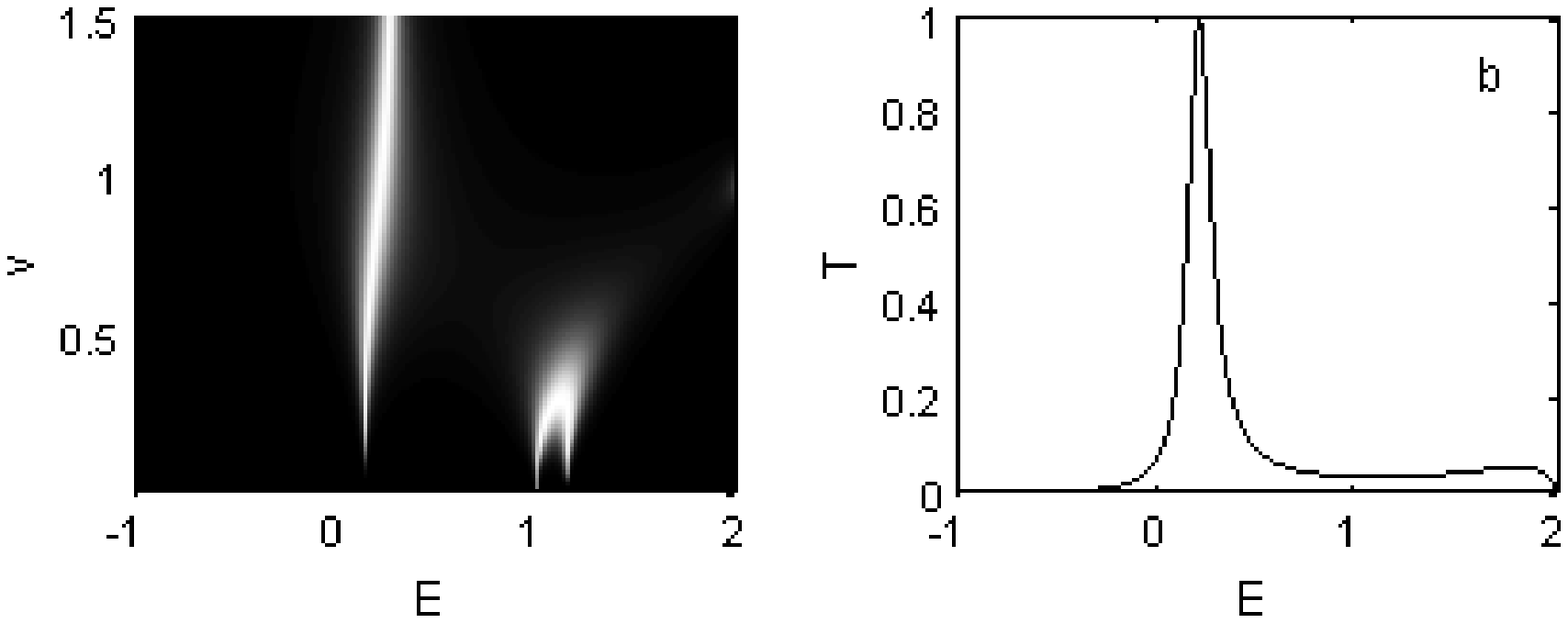}
\caption{(a) The probability $T$ for transmission through the double QD
versus $E$ and $v$ for $L_c=8.5355$. (b) The transmission
probability as a function of $E$ for fixed $v=0.85$.
It has one narrow peak on the background caused by the two broad resonance
states. Parameters: $\varepsilon_1=1, ~\epsilon(L)=2-L/5, ~u=1/4$
 as in  Fig. \ref{fig1}. }
\label{fig6}
\end{figure}

Up to now, we traced the appearance of a branch point in the complex
plane by enlarging the coupling strength $v$ between system and leads. In such
a case, the branch points at which two eigenvalues coalesce,
appear in a natural manner. It is less evident that the branch points in the
complex plane can be seen in all  parameters of the double QD system
that define Eq. (\ref{branch}).
We can take  arbitrary but fixed values of $v$ and $u$
and consider the length $L$ or even the energy $E$ as a parameter in
order to  trace the coalescence of $z_1$ and $z_3$ at $L_c$ and $E_c$.
The corresponding equations for achieving the
coalescence are
\begin{eqnarray}
\label{LcEc}
\epsilon(L_c)=\varepsilon_1^c\pm\sqrt{v_c^4-8u_c^2} \; ; \qquad
E_c=\pm \; \frac{2}{v_c^2}\; \sqrt{v_c^4-8u_c^2} \, .
\end{eqnarray}
A whole branch cut occurs along  $L$ when $u=u_c$, $v=v_c$ and $E=E_c$
are fixed but $\varepsilon_1$ is not fixed.
We consider in the following  one branch point corresponding to a fixed
value of $\varepsilon_1$.

The  case with $L$ as a parameter is shown in Fig. \ref{fig5}  for the
same  double QD system as in Fig. \ref{fig1}, but $v=1$.
There are two branch points in the complex plane
corresponding to  $E_{1c}=\sqrt{2}, ~L_{1c}=1.4645$ and $E_{2c}=
-\sqrt{2}, ~L_{2c}=8.5355$.
When $L<L_{1c}$ and $E> \sqrt{2}$, the two levels 1 and 3
avoid crossing as in the  foregoing cases.
In the region $L_{1c} < L < L_{2c}$ and $-\sqrt{2}<E<\sqrt{2}$,
the levels do not cross at all in the
complex plane due to their  different widths: one of them is trapped by the
other one due to the strong interaction via
the continuum (i.e. via the modes propagating in the leads). For $L>L_{2c}$
and $E< -\sqrt{2}$, the levels
again avoid crossing in the complex plane since the widths
and with them the external coupling of the states via the continuum decrease
in approaching $E= -2$.

The appearance of  two branch points in the complex plane in Fig. \ref{fig5}
illustrates in a very convincing manner the interplay between
internal and external interaction in approaching a branch point.
In any case, a branch point  separates regions with avoided level crossing
($L<L_{1c}, ~L>L_{2c}$) from those  without any crossing of the levels
($L_{1c} < L < L_{2c}$) in the complex plane.
One should underline, however, that the first branch point influences
the physical observables such as the transmission probability
[Fig. \ref{fig6} (a)], indeed. The second branch point occurs as a threshold
effect far
from the energies $E_1$ and $E_3$ of the two states. The two eigenvalues
$z_1$ and $z_3$ coalesce at the energy $E_c=-\sqrt{2}\ll E_k-\Gamma_k/2,E_l
-\Gamma_l/2$, i.e. at the tails of the resonance states.
This does not have any influence on the transmission probability.

In Fig. \ref{fig6} (b), the transmission probability is shown at $L=L_{2c}$.
It shows one peak, caused by the narrow resonance state, on the background
created by the two broad resonance states. The narrow resonance is of
Fano type by
taking into account that the background decreases in approaching the two
borders $E=\pm2$. The transmission probability for other values of
$L>L_{1c}$ is similar to that shown in Fig \ref{fig6}.

\begin{figure}[t]
\includegraphics[width=.6\textwidth]{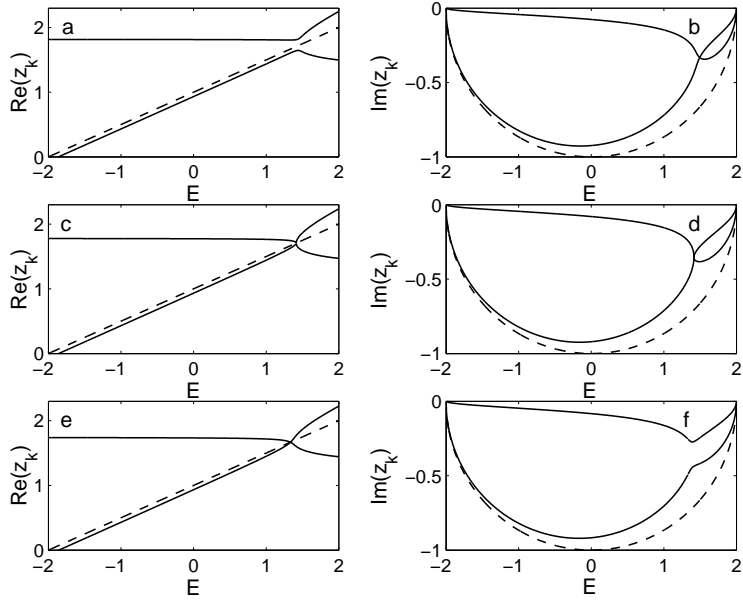}
\caption{The evolution of the real (left column)
and imaginary (right column) parts of the eigenvalues $z_k$,
~$k=1, 3$ (solid lines), ~$k=2$ (dashed line),
as a function of energy for transmission through the same
double QD system as in Fig. \ref{fig1}, but $v=1$ as in Fig. \ref{fig5}.
The  point of coalesced eigenvalues  is $E_c=\sqrt{2}$.
The critical length is $L_c=1.4645$.
(a,b) $L=L_c-0.1$, (c,d) $L=L_c$, and (e,f) $L=L_c+0.1$. }
\label{fig7}
\end{figure}

In Fig. \ref{fig7}, we show the analogue pictures for the $E$ dependence of
the eigenvalues $z_k$. Due to the fact that the energy is bounded from below
($E=-2$) and above ($E=2$), the energy dependence of $Im(z_k)$ can not be
neglected. It is especially large for states that are strongly coupled to the
continuum.  While the energy dependence of  $Im(z_k)$ is more or less
symmetrically around $E=0$, the  $Re(z_k)$  show an unsymmetrical
behavior as a function of  energy. It is of special interest, that
the branch points in the complex plane appear also
in the energy dependence of $Re(z_k)$ and $Im(z_k)$.
An example is the branch point at $E_c=\sqrt{2}, ~L_c=1.4645$
that can be seen in Fig. \ref{fig7}.

\section{Transmission through a double dot system with different coupling
strengths to the two leads}

Till now we considered the case that the double QD is coupled to the
left and to the right reservoir with the same strength  $v$. The
couplings may be, however,  different from one another.
Such a case is interesting, also
from a theoretical point of view, since the
effective Hamiltonian becomes  unseparable when the two coupling strengths
differ from one another. This is  in contrast to (\ref{Heff3}) where
the double QD is assumed to be coupled symmetrically  to the reservoirs
and, according to (\ref{poles3}) and (\ref{psiHeff3}),
the eigenstate $|2)$ does not interfere with the other two states
$|1)$  and $|3)$.

Following  \cite{1} we can write (\ref{Heffgen}) as follows
\begin{eqnarray}\label{Heffvw}
 \langle  m|H_{\rm eff}|n\rangle & =& E_m\delta_{mn}+
\sum_{C=L,R}\frac{1}{2\pi}\int_{-2}^2 dE'
\frac{V_m(E',C)V_n(E',C)}{E+i0-E'}\nonumber\\ &=&
E_m\delta_{mn}-\Big(v^2\psi_m(1)\psi_n(1)-w^2\psi_m(3)\psi_n(3)\Big)e^{ik},
\end{eqnarray}
where $v, w$ are the coupling strengths between the system and,
respectively, the  right and left reservoirs.
Substituting the eigenstates of the closed
double QD system (\ref{HB3}) into (\ref{Heffvw}) we obtain the
following expression for the (symmetrical) effective Hamiltonian
\begin{equation}
\label{Heff3vw} H_{\rm eff}=\left(\begin{array}{ccc}
E_1^B-\frac{(v^2+w^2)u^2e^{ik}}{2\eta(\eta+\Delta\varepsilon)} &
-\frac{u(v^2-w^2)e^{ik}}{2\sqrt{\eta(\eta+\Delta\varepsilon)}} &
\frac{u(v^2+w^2)e^{ik}}{2\sqrt{2}\eta} \cr
-\frac{u(v^2-w^2)e^{ik}}{2\sqrt{\eta(\eta+\Delta\varepsilon)}} &
\varepsilon_1-(v^2+w^2)e^{ik}/2 &
\frac{u(v^2-w^2)e^{ik}}{2\sqrt{\eta(\eta-\Delta\varepsilon)}}  \cr
\frac{u(v^2+w^2)e^{ik}}{2\sqrt{2}\eta} &
\frac{u(v^2-w^2)e^{ik}}{2\sqrt{\eta(\eta-\Delta\varepsilon)}} &
E_3^B-\frac{(v^2+w^2)u^2e^{ik}}{2\eta(\eta-\Delta\varepsilon)}\cr
             \end{array}\right) \; .
\end{equation}

\begin{figure}[t]
\includegraphics[width=.6\textwidth]{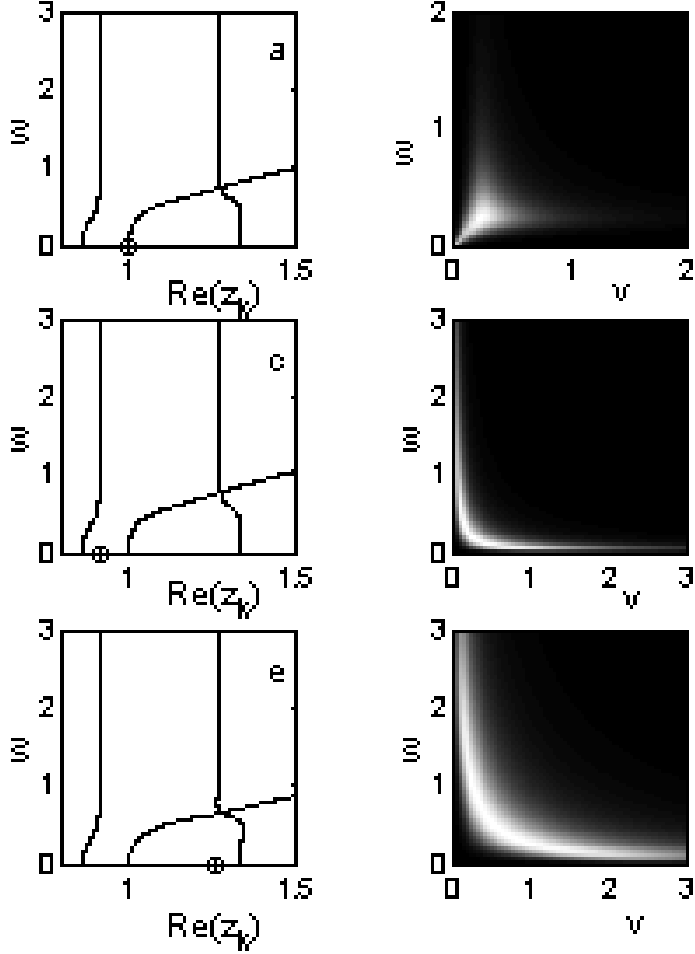}
\caption{Left column: The evolution of the real parts of the eigenvalues
of (\ref{Heffvw})  as a function of $w$ for $v=0.1, ~E=1.0$ (a), ~$v=0.06,
~E=0.92$  (c),
and $v=0.1, ~E=1.26$ (e). The parameters of the closed double
QD system are  $L=4, ~u=0.15, ~\varepsilon_1=1, ~\epsilon(L)=2-L/5$.
The circles at the x-axes denote the energies $E$.
Right column: The transmission probability through the double QD
versus coupling $v$ with the left reservoir and $w$ with the right reservoir.
The energies $E$ are the same as in the corresponding figures
of the left column.
}
\label{fig8}
\end{figure}

The transmission probability for  a system with different couplings of the
double QD to the reservoirs  demonstrates new
features that appear when $v$ and $w$ differ strongly from one another
(Fig. \ref{fig8}). In the calculations, we have chosen the following
parameters for the double QD system: $\epsilon(L)=2-L/5,\; L=4,\; u=0.15, \;
\varepsilon_1=1$. Then from (\ref{ED3}) we have $E_1^B=0.8665, \;
E_2^B=\varepsilon_1=1, \; E_3^B=1.3345$ for the three states of the closed
system.  The positions of the real parts $Re(z_k), k = 1, 2, 3,$
of the three eigenvalues of the effective Hamiltonian $H_{\rm eff}$
are given in Fig. \ref{fig8}, left column, for $E=1.0, ~0.92$ and 1.26.

Let us at first tune the energy of the incident particle to be
resonant with the eigenenergy $E=E_2^B=1$
of the closed system. As it can be seen from Fig. \ref{fig8} (a),
we can have resonant transmission through the system at this energy
only for  $w<1/2$. Correspondingly, the transmission probability decreases
for large $w$, Fig. \ref{fig8} (b). Next, let us take $E=0.92$
that approaches $E_1$ for  $w\approx 1/3$ according to  Fig. \ref{fig8} (c).
Resonance transmission through the system is possible,
at this energy,  only  when $w \ge 1/3$ and $v=0.06$. Since $Re(z_1)$ is
almost constant as a function of $w$ when $w > 1/3$, also the transmission
remains almost constant for $w>1/3$.
Obviously the transmission is symmetrical relative to $v\leftrightarrow w$.
As a result we obtain the peculiar picture of
transmission probability shown in Fig. \ref{fig8} (d). A similar
picture is obtained if the energy is tuned to the third eigenenergy
that is $E=E_3=1.26$ for large $w$, as shown in Figs. \ref{fig8} (e, f).
We mention, however, that at larger $u$ the transmission picture
is less peculiar.  Maximum transmission
appears when $w \approx v$ and $v$ is about 2 or 3 times larger than
$u$.

\section{Transmission through a double dot system with more than three
states}

We show now results of some calculations
for the transmission through a  more realistic double QD system
with more than one state in each of the single QDs.
The number of propagating
modes in the leads as well as in the wire, connecting the two single QDs,
is restricted to one as in the foregoing calculations.

\begin{figure}[t]
\includegraphics[width=.7\textwidth]{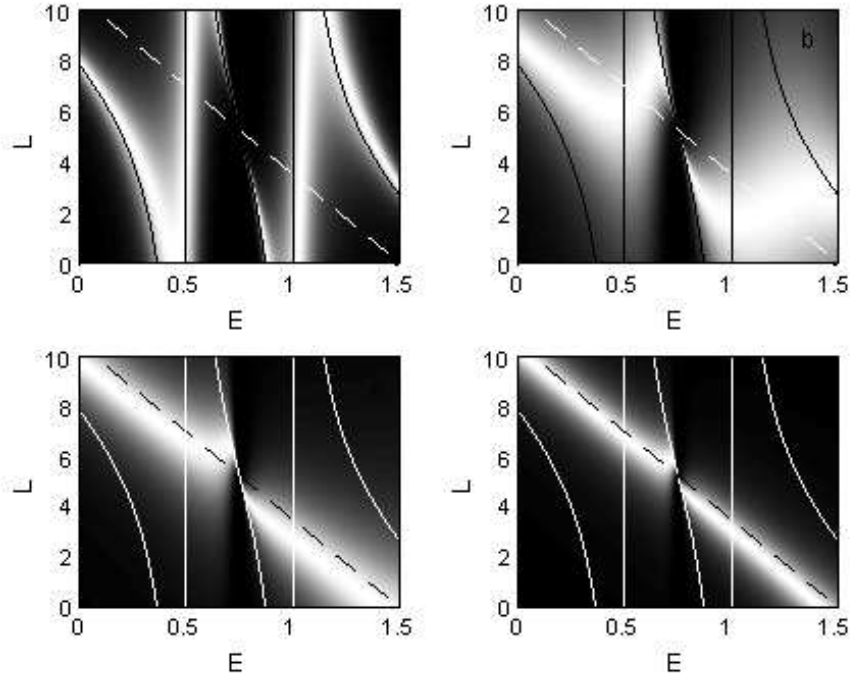}
\caption{ The transmission through a double QD versus $E$ and $L$
for $v=0.25$ (a), ~0.5 (b), ~0.75 (c) and 1.0 (d).
The solid lines represent the five real eigenvalues
$E_k^B$ of the Hamiltonian $H_B$ as a function of $L$. The dashed
lines show the eigenenergy of the wire
$\epsilon=3/2-L/7$. The eigenenergies of the two single QDs are
equal: $\varepsilon_1=1/2, ~\varepsilon_2=1$, and $u=0.25$. The
transmission zero at $E_0=3/4$ is independent of $L$ and $v$.
\label{fig9}}
\end{figure}

In Fig. \ref{fig9}, we show the transmission through such a double
QD system with two states in each single QD as a function of
energy $E$ and length $L$ for $u=0.25$
and for four different coupling strengths $v\leq 1$. The results
show the change of the transmission picture as a function of $L$
for different $v$. At small $v$, the transmission takes place
mainly at the energies  $E_k^B$ of the discrete states of the
double QD.  This behavior is called usually resonant transmission.
At larger $v$, however, the
transmission peaks have nothing in common with the positions
$E_k^B$ of the eigenstates of $H_B$. Here, the energy and $L$
dependence of the transmission follows basically that of the wave
inside the wire, $\epsilon=3/2-L/7$. The transmission picture given
in Fig. \ref{fig9} corresponds to those shown in \cite{1}.
Transmission zeros appear for all $v$ at $E^{(0)}_{s}=
(\varepsilon_1^{s}+\varepsilon_2^{s})/2$ where $\varepsilon_k^{s}
~(k=1,2 \, ; ~s=l,r) $ ~are the eigenenergies of, respectively,
the left and right single QD. It is $E^{(0)}_{l}=E^{(0)}_{r}=
3/4$ in Fig. \ref{fig9}.

\begin{figure}[t]
\includegraphics[width=.7\textwidth]{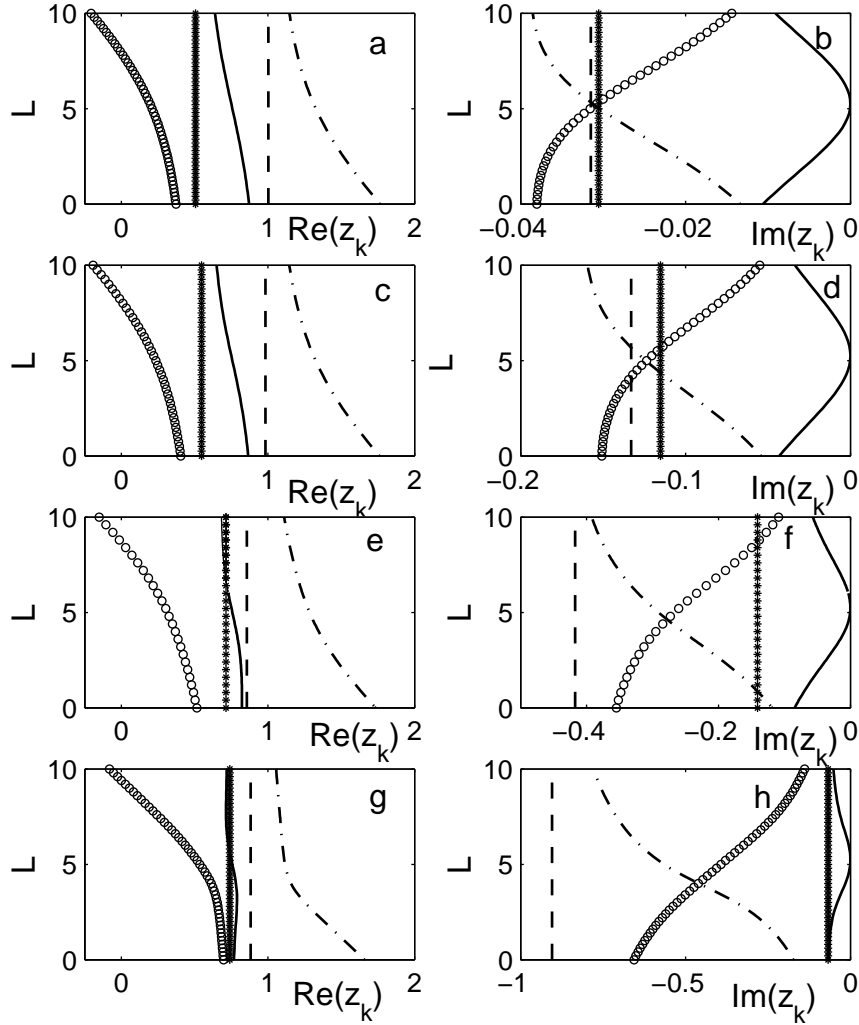}
\caption{
The evolution of real (left) and imaginary (right) parts of the
five eigenvalues of the Hamiltonian $H_{\rm eff}$  as a function of the
length $L$ for a double QD system. The  coupling of the
system to the continuum is $v=0.35$ (a, b), ~0.8 (c, d), and ~1.1 (e, f).
The parameters of the system are
$u=0.25$, ~$E=0.25$,  ~$\epsilon=3/2-L/7$.
The energies of the two single QDs are the same:
$\varepsilon_1=1/2, ~\varepsilon_2=1.$
The transmission of this double QD is shown in Fig.
\ref{fig9}
} \label{fig10}
\end{figure}

The eigenvalue pictures corresponding to Fig. \ref{fig9} are shown
in Fig. \ref{fig10}. As long as $v$ is small, the energies
$Re(z_k)$ show a  dependence on the parameter $L$ that is typical
for interacting (discrete) states. The $Re(z_{k,l})$ of the two
outermost states avoid crossing  at a certain $L=L^{\rm cr}$ where
the decay widths $2\, {Im}(z_{k,l})$ cross. At larger $v$, however, the
eigenvalue pictures change since the widths of the two outermost
states do no longer cross in the complex plane. Though  the
trajectories projected onto the energy axis cross at a certain
value of $L$,  the decay widths do not cross at all. This
is due to the large difference between $Im(z_1)$ and $Im(z_3)$
as a consequence of  resonance trapping (width bifurcation).

We can see from the eigenvalue trajectories Fig. \ref{fig10}
that the  picture \ref{fig9} (d) corresponds also to resonant transmission
in spite of the fact that its structure is completely different from that in
\ref{fig9} (a). The point is that the eigenvalues of $H_{\rm eff}$ differ
fundamentally from those of $H_B$ if  the coupling of the states via the
continuum is strong. The transmission peak appears at the position
of a narrow resonance state. Besides this state, there are two broad and
two narrow resonance states lying each very close to one another.
The interferences between them are obviously destructive.

Another interesting result seen in Fig. \ref{fig10} is that the decay
width of the state in the middle of the
spectrum vanishes at $L\approx 3$ for all
$v$. At this value of $L$, the middle state crosses the energy
$E^{(0)} = 3/4$ where the transmission is zero. For a discussion
of the transmission zeros see \cite{1}.

\begin{figure}[t]
\includegraphics[width=.8\textwidth]{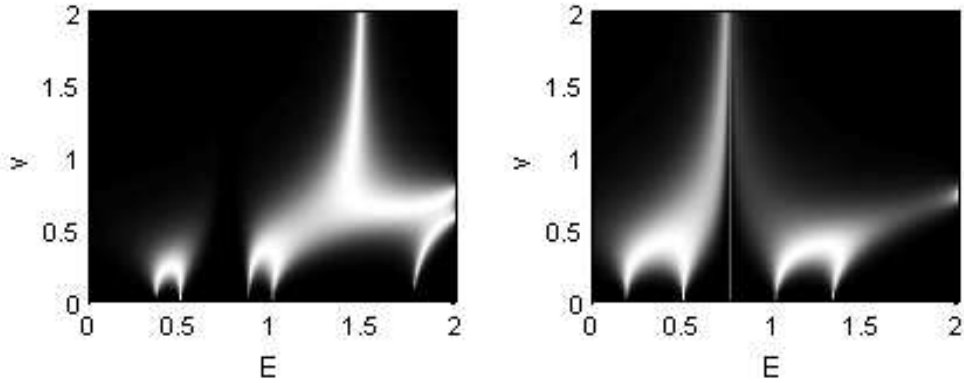}
\caption{
The transmission through a double QD versus $v$ and $E$
with the length $L=2$ (a) and 5 (b). The  parameters are
$u=0.25$,
~$\epsilon(L)=2-L/4, ~\varepsilon_1=1/2, ~\varepsilon_2=1$.
The transmission zero at $E_0=3/4$ is independent of $v$ and $L$.}
\label{fig11}
\end{figure}

In Fig. \ref{fig11}, the transmission through a double QD with altogether
five states is shown as a function of
energy and $v$  for  two different lengths of the wire,
$L=2 $ and 5. Each of the
two single QDs has two levels at $\varepsilon_1 =1/2$ and
$\varepsilon_2 = 1$, and the mode in the wire is $\epsilon(L) = 2-L/4$.
Transmission zeros appear at $E= 3/4$
(for a detailed discussion of the transmission zeros see \cite{1}).

\begin{figure}[t]
\includegraphics[width=.7\textwidth]{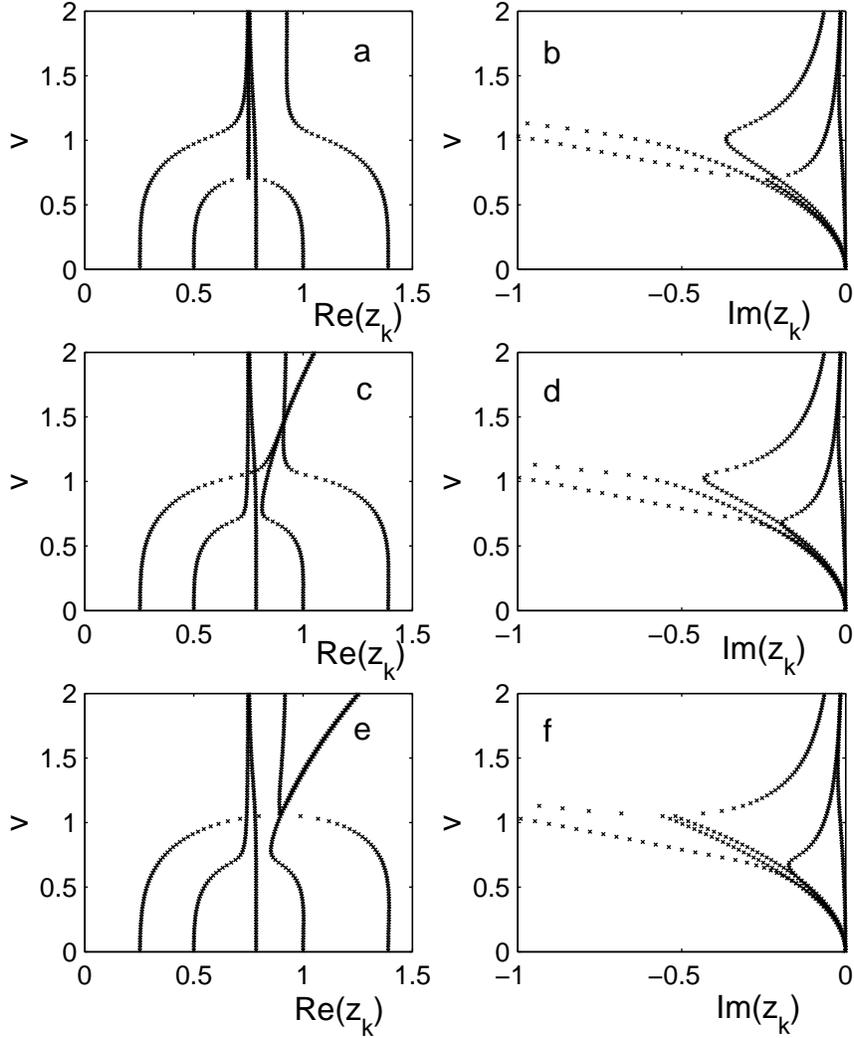}
\caption{The evolution of real (left) and imaginary (right) parts of the
five eigenvalues of the Hamiltonian $H_{\rm eff}$
as a function of the coupling strength
$v$  for a double QD. The length of the wire
is $L=0.7 $ (a, b), ~2 (c, d), and 3.03 (e,f).
Parameters:  $u=0.25$, ~$E=0.75$,
~$\epsilon(L)=2-L/4, ~\varepsilon_1=1/2, ~\varepsilon_2=1$.
The transmission of this double QD is shown
in Fig. \ref{fig11}.} \label{fig12}
\end{figure}

The eigenvalue pictures corresponding to Fig. \ref{fig11} at $E=0.75$
are shown in Fig. \ref{fig12}.
We see a bifurcation of the widths as discussed in Sect. III
as well as the corresponding branch points in the complex plane.
At large $v$, there are two broad resonance states according to the two modes
propagating in the two leads. The remaining three states are
narrow at large $v$. They are trapped by the two broad states. As shown
in Fig. \ref{fig12}, the two outermost states coalesce only at $L=3.03$.
The resonance state in the middle of the spectrum
coalesces, however,  with  another state at lower energy
for all three lengths $L$ shown in Fig. \ref{fig12}.

The eigenvalue pictures calculated at different energies
differ from one another in some details.
The eigenvalue picture \ref{fig12} corresponds to Fig. \ref{fig1} calculated
at a positive energy $E$. The two broad states  are shifted to higher energy
when $v$ is large. The shift is in the opposite direction when the eigenvalue
pictures are calculated at negative energy. The calculation at $E=0$ gives a
symmetrical picture  corresponding to Fig. \ref{fig3}.
In this case, the positions  of all states at large $v$ are almost constant.
The resonance trapping mechanism occurs symmetrically  at $E=0$:
the two outermost states coalesce at a somewhat higher value of $v$ than
the two states lying nearer to the center of the spectrum.  The state
in the middle of the spectrum does not coalesce with any other state.
It corresponds
to the mode moving in the wire and is symmetrically coupled to
the states at higher and at lower energy when $E=0$. This
result corresponds completely to those shown in Figs. \ref{fig3}.

The figures show clearly that the transmission peaks appear at the
positions of the eigenstates of $H_B$ only when $v$ is small. At
larger $v$, the transmission is determined by interferences
between the contributions from the different states. Nevertheless, it is
resonant  in relation to the eigenstates of the effective Hamiltonian
$H_{\rm eff}$.  Level repulsion at small $v$ and level
attraction at large $v$ cause  features of the transmission
pictures for a double QD with altogether five states
(Figs. \ref{fig11} and \ref{fig12}) that are the same as those of a double QD
with altogether only three states (Figs. \ref{fig1} to \ref{fig4}). The only
difference is the appearance of transmission zeros
(Fig. \ref{fig11}) when the two single QDs
are coupled to one another so that the double QD is effectively different from
a 1d-chain as in Figs. \ref{fig11} and \ref{fig12}, see \cite{1}.

\begin{figure}[t]
\includegraphics[width=.6\textwidth]{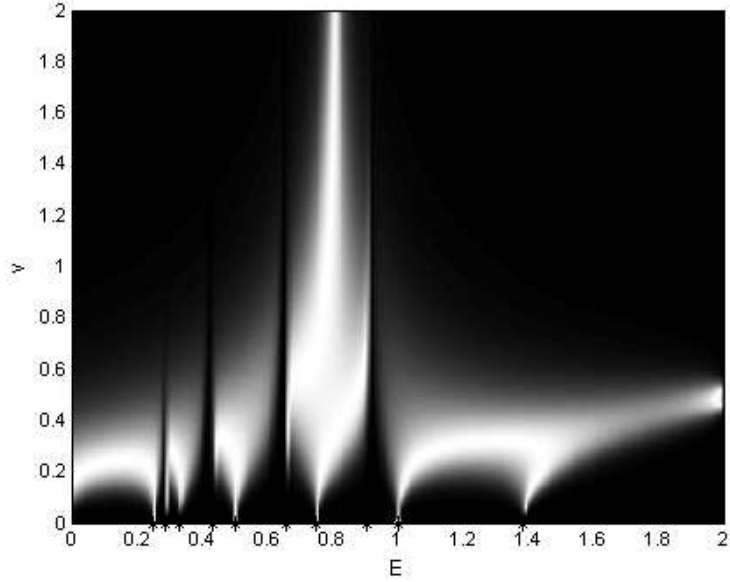}
\caption{ The transmission through a double QD  versus $v$ and $E$
with the parameters
$L=1.5$ and $u=0.2$. Each single QD has five levels at $\varepsilon_i =1/4,
~1/3, ~1/2, ~3/4, ~1$. The energy in the wire is $\epsilon
=1-L/8$. The four transmission zeros are independent of $v$ and $L$.}
\label{fig13}
\end{figure}

\begin{figure}[t]
\includegraphics[width=.6\textwidth]{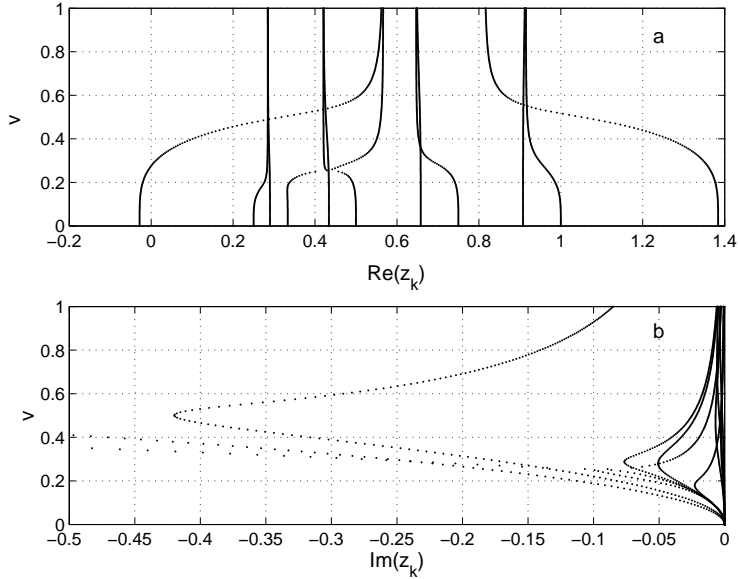}
\caption{The evolution of the 11 eigenvalues $z_k$ of the effective
Hamiltonian $H_{\rm eff}$ as a function of $v$ at $E=0$. (a) $Re(z_k)$,
(b) $Im(z_k)$. $L=1.5, ~u=0.2$.  Each single QD has five levels at
$\varepsilon_i =1/4, ~1/3, ~1/2, ~3/4, ~1$.
The eigenenergy of the wire is $\epsilon =1-L/8$.
The transmission of this double QD is shown in Fig. \ref{fig13}.
}
\label{fig14}
\end{figure}

In Fig. \ref{fig13}, the transmission through a QD with
five states in each single QD  is shown, and  Fig. \ref{fig14} gives
the corresponding eigenvalue trajectories of all 11 states. The
main features discussed for the cases with a smaller number of
states remain.
This holds true also for the transmission zeros the positions of
which are determined by the energies of the eigenstates of the two
single QDs. One of the differences to the cases with altogether three
or five states is the following. The eigenenergy trajectories
at $E=0$ are symmetrical around the energy $E=0$ in
Fig. \ref{fig3} with only one state in each single QD, while the
symmetry is somewhat disturbed in Fig. \ref{fig14} with more states
in each single QD. In the latter case, the two outermost states do
not approach each other completely. The lower state approaches one
of the  states out of the middle, and the upper state becomes
trapped by these two states. As a consequence, the region with
maximum transmission does not occur in the middle of the spectrum
but at a somewhat lower energy. The reason for this asymmetry is
the following: the functions $Re (z_k)$  of ten states are
raising with energy while all the $Im (z_k)$  are vanishing
at the two limits $E=\pm 2$ of the energy window (compare Fig.
\ref{fig7}). Therefore, the widths of the states at lower energy
are larger than those of the states at higher energy so that they
trap the higher-lying states. For details of the resonance
trapping phenomenon see \cite{rep}.

Common to all the  pictures  shown in this section is that the
single-channel transmission  through a double QD is of resonant
character although its structure depends strongly on the strength
$v$ by which the dot is coupled to the attached leads.
The point is that the evolution of the eigenvalues
of the effective Hamiltonian $H_{\rm eff}$ as a function of
external parameters changes fundamentally at branch points in the
complex plane.
The transmission through the double QD shows a correspondingly
sensitive dependence on the external parameters.
Qualitative changes  in the transmission picture are
caused by  branch points in the complex plane which separate
the scenario with avoided level crossing  from that
without any crossing in the complex plane.
While  transmission  occurs in the whole energy region with several peaks
in the  case with avoided level crossings, there is a smaller number
of peaks of mostly different height  in the case
without any level crossings in the complex plane.
The position of these peaks changes as a function of $L$.
Common to both scenarios are only the $L$ independent transmission zeros
(for a detailled discussion of the transmission zeros  see \cite{1}).

The two coupling strengths $v$ and $u$ stand, respectively,
for the coupling of the double QD as a whole to the leads
(environment) and the coupling of the two single QDs to the wire (inside the
double QD system).
The ratio $v/u$ characterizes therefore the ratio between external and
internal interaction of the states
of an open quantum system. When the
external coupling is  much larger than the internal coupling,
the external coupling of the levels via the modes propagating in the two leads,
prevents the formation of a uniform QD. In the opposite case of
large internal coupling, the relatively weak external coupling is unable
to break the  uniform QD. Most interesting is, of course,
the transition region between the two different types of bonds
in double QDs.

It is worthwhile to notice  the following. The two
levels that are the outermost ones of the spectrum, cross or avoid crossing
in the complex plane at $E=0$. The distance in energy to the  crossing or
avoided crossing, that occurs between  two other levels,
is smaller than their decay widths.
That means, effectively all states are involved in the scenario
of avoided level crossing  in the complex plane.

Additionally, we mention that the dependence of the transmission
on the length $L$  of the wire
is determined by the manner the wave propagates inside the wire. It can be
replaced  by another relation between $\epsilon$ and $L$ than
that   used in our  calculations or by  the analogue relation between
$\epsilon$ and the width $d$ of the wire. In the last case, $L$ can be kept
constant in studying the dependence of the transmission from $d$, see
the discussion at the end of Ref. \cite{1}.


\section{Summary}

The results considered in the present paper are obtained in the
formalism worked out in \cite{1} for the description of a double QD system.
The formalism is based on the $S$ matrix theory with use of the
effective Hamiltonian that describes the spectroscopic
properties of the open quantum system.
The formalism is applied in \cite{1} to the description of
transmission zeros  in the conductance through double QDs. These zeros
are determined by the spectroscopic properties of the constituents of the
double dot system and by the manner the single QDs are coupled.
They appear  at all ratios $v/u$ of the coupling strengths.
Our present study is devoted, above all, to the transmission peaks.
Their positions and widths depend on the ratio $v/u$ and are
influenced by branch points in the complex plane.
At these points, the
transition between the two scenarios with avoided level crossing
and no crossing in the complex plane takes place. In any case, the
transmission is resonant.

As long as $v/u$ is small, the levels repel in energy
(as the discrete eigenstates of $H_B$) and the decay widths
of the different states are  of comparable value. This causes some
spreading of the transmission probability over a
relatively large energy region.  At large $v/u$, however, the
levels attract in energy and the decay widths bifurcate. This causes
transmission peaks at the positions of the narrow states that appear
on the smooth background created by the broad states. The positions of the
transmission peaks depend, in this case, strongly
on the length of the wire or
on another parameter that controls the propagation of the mode
inside the wire.
The two different scenarios are separated by a branch point in the complex
plane. At such a point,
two eigenvalues ($z_k$ and $z_l$) of the effective Hamiltonian
coalesce at the energy $E=E_c$. Sometimes,
$E_c=E_k = E_l$. Mostly however  $E_k \ne Re (z_k)|_{E=E_c}$ and $E_l \ne
Re ( z_l)|_{E=E_c}$,
and the branch point in the complex plane is not a double pole
of the $S$ matrix.

We underline that the resonance phenomena
appearing in the transmission through double QDs are the same as those
observed in, e.g., the scattering on nuclei or atoms \cite{rep}.
The role of the branch points in the complex plane for
the transmission through a double dot system  agrees with that discussed
in a  schematical study
\cite{ro01}  and for a double-well system \cite{vanroose}.
In our model double QD, however, the energy dependence of the eigenvalues
$z_k$ of the effective Hamiltonian $H_{\rm eff}$ is relatively strong.
Especially  $Im(z_k)$ shows a strong energy dependence
due to the energy window
with  thresholds at a lower and an upper finite energy.
The spectrum is therefore bounded  from below and from above,
and   the eigenvalues  of the effective Hamiltonian
cannot satisfyingly be approximated  by the poles of the $S$ matrix.

The results discussed here are true for single-channel transmission
through a double QD system that consists of
two single QDs with similar energy spectra and  a narrow wire
that couples the two single QDs and allows the
propagation of only one mode.  When the energy spectra
of the two single QDs are very different from one another and the
coupling strength
$u$ to the wire is small, the transmission picture at large $v$ differs from
that discussed above. In such a case, the transmission is hindered at large
$v$, above all due to the energy gap between the levels of the two
single QDs through which the transmission takes place.

In the present paper, the behavior of a simple model
is considered that reflects many characteristic features of realistic
double QDs
with more complicated structure, see \cite{1}. The results obtained may guide
the construction of double QDs. The position of transmission zeros
and transmission peaks can be controlled by varying the coupling strengths
$v$ and $u$ as well as the propagation of the mode inside the wire.
An example is the broad plateau with maximal transmission shown in
Fig. \ref{fig4} (b). Using the interplay between
internal and external interaction allows to control the properties of QDs
in a systematic manner.

\acknowledgments  We thank Erich Runge for critical reading the
manuscript. A.F.S. thanks  the Max-Planck-Institut
f\"ur Physik komplexer Systeme for hospitality.
This work has been supported by the RFBR grant
04-02-16408.
\\

$^{*}$ e-mails: rotter$@$mpipks-dresden.mpg.de;
almsa$@$ifm.liu.se, almas$@$tnp.krasn.ru,
almsa$@$mpipks-dresden.mpg.de


\begin{thebibliography}{99}

\bibitem{kato} T. Kato, {\it Perturbation Theory of Linear Operators}
(Springer, Berlin 1966).

\bibitem{moisrep} N. Moiseyev, Phys. Rep. {\bf 302}, 211 (1998).

\bibitem{ro01} I. Rotter, Phys. Rev. E {\bf 64}, 036213 (2001).

\bibitem{dicaro} F.M. Dittes, W. Cassing and I. Rotter, Z. Phys. A {\bf 337},
243 (1990);
I. Rotter, Rep. Prog. Phys. {\bf 54}, 635 (1991).

\bibitem{mudiisro}
M. M\"uller, F.M. Dittes, W. Iskra, and I. Rotter
Phys. Rev. E {\bf 52}, 5961 (1995).

\bibitem{marost} A.I. Magunov, I. Rotter,
and S.I. Strakhova, J. Phys. B: At. Mol. Opt. Phys. {\bf 32}, 1669 (1999);
J. Phys. B: At. Mol. Opt. Phys. {\bf 34}, 29 (2001).

\bibitem{rep} J. Oko{\l}owicz, M. P{\l}oszajczak, and I. Rotter,
Phys. Rep. {\bf 374}, 271 (2003).

\bibitem{mois} E. Narevicius and N. Moiseyev, Phys. Rev. Lett. {\bf 84}, 1681
(2000); J. Chem. Phys. {\bf 113}, 6088 (2000).

\bibitem{newton} R.G. Newton, {\it Scattering Theory of Waves and Particles},
Springer, New York, 1982.

\bibitem{atom1} O. Latinne, N.J. Kylstra, M. D\"orr, J. Purvis,
M. Terao-Dunseath, C.J. Joachain, P.G. Burke, and C.J. Noble,
Phys. Rev. Lett. {\bf 74}, 46 (1995).

\bibitem{atom2}
N.J. Kylstra and  C.J. Joachain, Europhys. Lett. {\bf 36}, 657 (1996);
Phys. Rev. A {\bf 57}, 412 (1998).

\bibitem{atom3}
W. Vanroose, P. Van Leuven, F. Arickx, and J. Broeckhove,  J. Phys. A:
Math Gen. {\bf 30}, 5543 (1997).

\bibitem{vanroose} W. Vanroose, Phys. Rev. A {\bf 64}, 062708 (2001).

\bibitem{thomas} T. Meier, A. Schulze, P. Thomas, H. Vaupel, and K. Maschke,
Phys. Rev. B {\bf 51}, 13977 (1995).

\bibitem{marost4} A.I. Magunov, I. Rotter, and S.I. Strakhova,
Phys. Rev. B {\bf 68}, 245305 (2003).

\bibitem{1} I. Rotter and A.F. Sadreev, cond-mat/0403184 (2004).

\bibitem{qdot1}
W.G. van der Wiel, S. De Franceschi, J.M. Elzerman, T. Fujisawa,
S. Tarucha, and L.P. Kouwenhoven, Rev. Mod. Phys. {\bf 75}, 1 (2003).

\bibitem{qdot2}
J.P. Bird, R Akis, D.K. Ferry, A.P.S. de Moura, Y.-C. Lai, and
K.M. Indlekofer, Rep. Progr. Phys. {\bf 66}, 1 (2003).

\bibitem{saro}
A.F. Sadreev and I. Rotter, J. Phys. A: Math. Gen. {\bf 36}, 11413
(2003).

\bibitem{korsch} F. Keck, H.J. Korsch, and S. Mossmann, J. Phys. A: Math. Gen.
{\bf 36}, 2125 (2003).


\end{thebibliography}
\end{document}